\documentstyle[12pt]{article}

\newcommand{\be}{\begin{equation}}
\newcommand{\ee}{\end{equation}}
\newcommand{\no}[1]{:\!#1\!:}
\newcommand{\bz}{{\bf Z}}

\begin{document}

\begin{titlepage}

\begin{LARGE}
  \vspace{1cm}
  \begin{center}
     {Perturbative BPS-algebras in superstring theory.}
  \end{center}
\end{LARGE}

\vspace{15mm}

\begin{center}

  C.D.D. Neumann

  \vspace{2mm}

  {\it Department of Mathematics}\\
  {\it University of Amsterdam, }\\
  {\it Plantage Muidergracht 24, Amsterdam}\\
  {\it Email: {\tt neumann@wins.uva.nl}}\\
  \vspace{40mm}

  \begin{large} ABSTRACT \end{large}

  \par

\end{center}
\begin{normalsize}
This paper investigates the algebraic structure that exists
on perturbative BPS-states in the superstring, compactified
on the product of a circle and a Calabi-Yau fourfold. This
structure was defined in a recent article by Harvey and Moore.
It shown that for a toroidal compactification this algebra
is related to a generalized Kac-Moody algebra. 
The BPS-algebra itself is not a Lie-algebra.
However, it turns out to be possible to construct
a Lie-algebra with the same graded dimensions, in terms of
a half-twisted model. The dimensions of these algebras are
related to the elliptic genus of the transverse part of the 
string algebra. Finally, the construction is applied to an
orbifold compactification of the superstring.
\end{normalsize}
\end{titlepage}
\eject

\section{Introduction.}
\setcounter{equation}{0}

This paper investigates the algebraic structure
that exists on perturbative BPS-states in the superstring,
and its relation to generalized Kac-Moody algebras (GKM's).
This algebra structure was defined by Harvey and Moore
in \cite{BPS}, where they showed that in the toroidal
compactification of the heterotic string, this BPS-algebra
is closely related to a GKM-algebra. This Lie-algebra actually
comes from the left-moving part of the BPS-vertex operators.

In this article I show that this structure also
exists on the left-moving part of a toroidal compactified
superstring. To this end, I start by reviewing the
construction of this left-moving part in the RNS-formulation,
which is best suited to compute OPE's, in terms of
vertex algebras. I carefully avoid the introduction
of operators with fractional spin (parafermions), by
splitting the algebra into a transverse part, and a part
that combines ghosts and lightcone coordinates.
This turns out to be useful for generalizations.
I show that after GSO-projection this algebra
becomes a vertex super algebra (combining all ghostpictures).
By a suitable choice of grading (ghostnumber minus
ghostpicture), Lian and Zuckerman's dot and bracket
products \cite{LiZu} define a homotopy Gerstenhaber
algebra on this space, and then in cohomology
a Lie-algebra. Modding out the ghostpicture-equivalence
gives a GKM-algebra which is graded by
the lightcone momentum lattice. 

The graded dimensions of this algebra can be
expressed in terms of the (chiral) elliptic genus
of the transverse algebra. The GKM-algebra is,
however, not graded by the $U(1)$ gradation
which is present in the transverse algebra
(so that these algebras cannot be identified
with algebras like those constructed by Gritsenko and
Nikulin \cite{grit}, which do seem to have such a gradation). 

Then the relation between this algebra and
the full BPS-algebra of the superstring, compactified
on $S_1 \times C_4$ for a Calabi-Yau fourfold $C_4$
is investigated. Under the assumption that the
right-moving momentum has no component in the $C_4$
direction, the algebra closes without the use
of Lorentz-boosts, as introduced by Harvey and Moore.
For the torus compactification,
the algebra structure is the tensor product
of a Lie-algebra on the left-movers and a Lie-algebra
on the right movers (and thus not a Lie-algebra).
The number of BPS-states can be counted essentially
by the elliptic genus of $C_4$. 

It is then pointed out that there does exist a
Lie-algebra on a space with the same graded
dimensions, defined in terms of a half-twisted
model \cite{halftwist}, where essentially the
algebra on the right moving groundstates is
replaced by the associative algebra on the 
chiral ring.

In the last section I look at these structures
in a somewhat more complex model, where the
transverse algebra is replaced by the
sigma-model of the torus orbifold $T_8/\bz_2$.
The chiral-orbifold resembles in many ways the
construction of the monster Lie algebra \cite{moon},
which can be viewed as the Lie algebra
on states of a chiral bosonic string
compactified on $T_2\times (T_{24}/\bz_2)$,
where $T_{24}$ is defined by the Leech-lattice.

\section{Vertex algebras.}
\setcounter{equation}{0}

In this section I give definitions of the types
of vertex algebras that I will use as building blocks
for the superstring algebras. They are vertex (super)-algebras,
(I will sometimes omit the word super), and
$\bz_2\times\bz_2$-graded vertex algebras,
which would be called $\bz_2^3$-graded vertex para-algebras
in the terminology of \cite{spinor}.
Next, I review a general method to construct such
algebras from lattices.

\subsection{Formal definitions.}

A vertex algebra is defined as follows \cite{FLM,spinor,vrtx}.
Let $V$ be a vector space over the complex numbers ${\bf C}$,
which is $\bz_2$-graded, $V=V^0\oplus V^1$. Elements in
$V^0$ will be called bosonic, elements in $V^1$ fermionic.
For every $v \in V$, introduce an operator valued function
$v(z)$, such that $(\lambda v + \mu w)(z)=\lambda v(z)
+\mu w(z)$ for all $v,w \in V$ and $\lambda,\mu \in {\bf C}$.
These vertex-operators may be viewed as generating functions
\be v(z) = \sum_{n\in\bz} \{v\}_n z^{-n-1} \ee
for the linear transformations $\{v\}_k: V \rightarrow V$. 
These transformations must satisfy
\be \{v\}_k w \in V^{a+b} \ee
for every $v\in V^a, w\in V^b$, and
$\{v\}_k w=0$ for $k$ sufficiently large.
The main identity that the operators must satisfy
is the so called Jacobi identity.
Let $C_i(r,c)$ be a circle in the $\zeta_i$-plane
of radius $r$ centered at $c$, and let $f$ be any polynomial in
$\zeta_1, \zeta_1^{-1}, \zeta_2, \zeta_2^{-1}, (\zeta_1-\zeta_2)^{-1}$.
Also let $v\in V^a, w\in V^b$. Then the Jacobi identity
states that for $0<r_1<r_2<r_3$ and $0 < \epsilon <
\min (r_3-r_2, r_2-r_1)$
\begin{eqnarray}
& & \oint_{ C_2(r_2,0) } \oint_{ C_1(r_3,0) }
    v(\zeta_1) w(\zeta_2) f d\zeta_1 d\zeta_2 \nonumber \\
& -(-1)^{ab} & \oint_{ C_2(r_2,0) } \oint_{ C_1(r_1,0) }
    w(\zeta_2) v(\zeta_1) f d\zeta_1 d\zeta_2 \\
& = & \oint_{ C_2(r_2,0) } \oint_{ C_1( \epsilon, \zeta_2) }
    (v(\zeta_1-\zeta_2)w)(\zeta_2) f d\zeta_1 d\zeta_2 \nonumber
\end{eqnarray}
This equality is to be interpreted as an equality
of matrix elements.

\vspace{1\baselineskip} \noindent
The definition of a $\bz_2\times\bz_2$ graded 
vertex algebra is a generalization of the above \cite{spinor}. 
In this case, $V$ is graded not only by $\bz_2$, but also
by $\Gamma=\bz_2\times\bz_2$, whose elements will be written
as $\{ 00,01,10,11 \}$. So
\be V^a = \bigoplus_{\gamma\in\Gamma} V^a_{\gamma} \ee
I will also use the notation $V_{\gamma}$ for
$V^0_{\gamma} \oplus V^1_{\gamma}$.
The vertex-operators $v(z)$ do no longer in general have an
expansion in integral powers of $z$. In fact, define
\be \Delta(\gamma_1,\gamma_2) = \left\{ \begin{array}{ll}
  0 & \mbox{if $\gamma_1=00$ or $\gamma_2=00$ or $\gamma_1=\gamma_2$} \\
  \frac{1}{2} & \mbox{otherwise} \end{array} \right. \ee
which is a bilinear symmetric function mod $\bz$.
Then if $v\in V^a_{\gamma_1}$ and $w\in V^b_{\gamma_2}$,
one has
\be v(z)w = \sum_{n\in\bz+\Delta(\gamma_1,\gamma_2)}
    \{v\}_n w z^{-n-1} \ee
\be \{v\}_n w \in V^{a+b+2\Delta(\gamma_1,\gamma_2)
    }_{\gamma_1+\gamma_2} \ee
The Jacobi identity must be modified as follows. Let
$v\in V^a_{\gamma_1}$, $w\in V^b_{\gamma_2}$ and $u\in V_{\gamma_3}$.
Furthermore, let
\be f(\zeta_1,\zeta_2) \in
  \zeta_1^{\Delta(\gamma_1,\gamma_3)}
  \zeta_2^{\Delta(\gamma_2,\gamma_3)}
  (\zeta_1-\zeta_2)^{\Delta(\gamma_1,\gamma_2)}
  {\bf C}[ \zeta_1,\zeta_1^{-1},
  \zeta_2,\zeta_2^{-1},(\zeta_1-\zeta_2)^{-1} ] \ee
Then under the same conditions for the contours
\begin{eqnarray}
& & \oint_{ C_2(r_2,0) } \oint_{ C_1(r_3,0) }
    v(\zeta_1) w(\zeta_2) u f d\zeta_1 d\zeta_2 \nonumber \\
& -\eta(\gamma_1,\gamma_2)(-1)^{ab} & \oint_{ C_2(r_2,0) } 
    \oint_{ C_1(r_1,0) }
    w(\zeta_2) v(\zeta_1) u f d\zeta_1 d\zeta_2 \\
& = & \oint_{ C_2(r_2,0) } \oint_{ C_1( \epsilon, \zeta_2) }
    (v(\zeta_1-\zeta_2)w)(\zeta_2) u f d\zeta_1 d\zeta_2 \nonumber
\end{eqnarray}
where $\eta(\gamma_1,\gamma_2)$ is a sign defined by the
following table (a choice $x=\pm 1$ must be made)
\be \begin{tabular}{c|cccc}
    $\eta$ & 00 & 01 & 10 & 11 \\ \hline
     00   & 1  & 1  & 1  & 1  \\
     01   & 1  & 1  & x  & -x \\
     10   & 1  & -x & 1  & x  \\
     11   & 1  & x  & -x & 1 \end{tabular} \ee
This ends the definition of the $\bz_2\times\bz_2$-graded vertex algebra.
Note that the subalgebras $V_{00}\oplus V_{01}$,
$V_{00}\oplus V_{10}$ and $V_{00}\oplus V_{11}$
all satisfy the definitions of a vertex super-algebra.

\vspace{1\baselineskip} \noindent 
The vertex algebras that I will consider all have
two distinguished vectors ${\bf 1},\omega\in V$
with the following properties.
${\bf 1}(z)$ is the identity operator on $V$. The
components of $\omega(z)$, defined by
\be \omega(z) = \sum_{n\in\bz} L_n z^{-n-2} \ee
(so $L_n=\{\omega\}_{n+1}$) generate the Virasoro algebra.
\be [L_m,L_n] = (m-n)L_{m+n} + \frac{m^3-m}{12} c \delta_{m+n} \ee
$c$ is called the rank of $V$. The operator $L_0$
defines a gradation of $V$
\be V = \bigoplus_n (V)_n, \mbox{\hspace{1cm}}
    (V)_n = \{v \in V \:|\: L_0 v= nv \} \ee
A vector $v \in (V)_n$ will be called homogeneous of 
(conformal) weight $n$, and for such a vector I use the
notation ${\rm wt}(v)=n$.
In this case
\be [ L_{-1}, v(z) ] = (d/dz) v(z) = \partial v(z) \ee
\be [ L_0, v(z) ] = \left[ z(d/dz) + {\rm wt}(v) \right] v(z) \ee
For the components of $v(z)$ this implies
\be [ L_0, \{v\}_k ] = ({\rm wt}(v)-k-1) \{v\}_k \ee 
This suggests the introduction of another notation for these components,
as follows
\be v_k = \{v\}_{ {\rm wt}(v)+k-1 } \ee
because one then has the following simple relation for
two homogeneous vectors $v,w \in V$
\be {\rm wt}( v_k w ) = {\rm wt}(w) - k \ee
The vertex operator $v(z)$ is now expanded as follows
\be v(z) = \sum_k v_k z^{-k-{\rm wt}(v)} \ee

\subsection{Lattice construction.}

A well known way to construct a vertex algebra is
the lattice construction \cite{FLM}. The input is
a lattice $\Lambda$ of dimension $n$. Let $e^1,\ldots,e^n$
be a $\bz$-basis for the lattice. Then the lattice
is characterized by its matrix of inner-products
$G^{ij}=\langle e^i,e^j\rangle$ (also known as Gramm-matrix
\cite{sloane}). I assume this matrix to be non-degenerate,
and also that $\langle\alpha,\alpha\rangle\in\bz$
for all $\alpha\in\Lambda$. (The polarization
formula then shows that the general inner product is
half integral).
Since $G$ is symmetric, there is a basis of the ${\bf R}$-span
of the $e^i$, call it $f^1,\ldots,f^n$, in which the
inner-product is diagonal, $\langle f^i,f^j\rangle=g^{ij}$
where $g=\mbox{diag}((-1)^k,(+1)^l)$, and $(k,l)$ is
the signature of the lattice.
Now introduce bosonic fields $\partial\alpha^i(z)$ where
$1\leq i\leq n$, with components defined by
\be \partial\alpha^i(z) = \sum_{k\in\bz} \alpha^i_k z^{-k-1} \ee
The components of these operators have commutation relations
\be [ \alpha^i_k, \alpha^j_l ] =
       k\langle f^i,f^j\rangle\delta_{k+l} \ee
Next, introduce invertible operators $\hat{e}^i$ where
$1\leq i\leq n$, satisfying
\be [\alpha^i_k,\hat{e}^j]=\delta_{k,0}
      \langle f^i,e^j\rangle \hat{e}^j \ee
\be \hat{e}^i \hat{e}^j =
  \kappa^{c(e^i,e^j)} \hat{e}^j \hat{e}^i \ee
where $\kappa$ is some primitive $4$-th root of unity, 
and $c(e^i,e^j)$ is a bilinear function from
$\Lambda\times\Lambda$ to $\bz_4$, which is antisymmetric,
and for which $c(\alpha,\alpha)=0$ for all $\alpha\in\Lambda$.
Now it is possible to define operators for every 
$\alpha\in\Lambda$ as follows. Let $\alpha=\sum g_ie^i
=\sum h_if^i$. Then
\be \partial\alpha(z) = \sum h_i\partial\alpha^i(z) \ee
\be \hat{e}_{\alpha} = (\hat{e}^1)^{g_1} \cdots (\hat{e}^n)^{g_n} \ee
Now since the two sets of operators $X=\{\hat{e}^i,\alpha_0^i\}$
and $Y=\{ \alpha_k^i \}$ (where $k\neq 0$) mutually commute,
a representation for the algebra can be written as a tensor
product of representations for $X$ and $Y$.
A representation for $X$ (the lattice algebra) can be constructed
by introducing a vacuum $|0\rangle$, and then for
every $\alpha\in\Lambda$ a vector
\be |\alpha\rangle = \hat{e}_{\alpha}|0\rangle \ee
Then the operators in $X$ act on this as
\be \alpha_0^i|\beta\rangle = \langle f^i,\beta\rangle|\beta\rangle \ee
\be \hat{e}_{\alpha}|\beta\rangle = e^{i\phi}|\alpha+\beta\rangle \ee
for some phase $\phi(\alpha,\beta)$. I will call this 
representation-space $\hat{\Lambda}$. The algebra $Y$ can viewed
as $n$ copies of a Heisenberg algebra.
A representation for this algebra can be constructed by
introducing a vacuum-vector ${\bf 1}$, and then requiring that
\be \alpha_k^i {\bf 1} = 0, \mbox{\hspace{1cm}for $k>0$} \ee
The representation space for one copy of the Heisenberg algebra
will be called $H$. Consequently, $V$ has the form
\be V = \hat{\Lambda} \otimes H^n \ee
and this space will be called $V(\Lambda)$. (In the sequel, I will
use this notation even if $\Lambda$ is not a lattice, but for
example a subset of points from a lattice).
A general vector in $V(\Lambda)$ has the form
\be \Psi=\alpha^{i_1}_{-k_1-1}\cdots
         \alpha^{i_m}_{-k_m-1}|\beta\rangle \ee
where the $k_i\geq 0$.
To it, one can associate the following vertex operator
\be \Psi(z)=e_{\beta}z^{\beta_0} \no{
    \left[\frac{d^{k_1}}{k_1!dz^{k_1}}\partial\alpha^{i_1}(z)\right]
    \cdots
    \left[\frac{d^{k_m}}{k_m!dz^{k_m}}\partial\alpha^{i_m}(z)\right]
    \exp\left( -\sum_{k\neq 0}\frac{\beta_kz^{-k}}{k}\right) } \ee
The normal ordering is defined in the usual way
\be \no{ \alpha^i_k \alpha^j_l } =
  \left\{ \begin{array}{ll} 
  \alpha^i_k \alpha^j_l & \mbox{if $k<l$} \\
  \alpha^j_l \alpha^i_k & \mbox{otherwise}
  \end{array} \right. \ee
The vertex operator can also be written symbolically as
\be \Psi(z)=\no{
    \left[\frac{d^{k_1}}{k_1!dz^{k_1}}\partial\alpha^{i_1}(z)\right]
    \cdots
    \left[\frac{d^{k_m}}{k_m!dz^{k_m}}\partial\alpha^{i_m}(z)\right]
    e^{\beta(z)} } \ee
which is more usual in string theory.
With these definitions, $V(\Lambda)$ becomes a vertex para-algebra,
if the cocycle $c(e^i,e^j)$ is chosen properly.
It is easy to define the $\bz_2$ grading in this algebra.
Namely, let
\be \Lambda^a = \{ \alpha\in\Lambda \:|\:
       \langle\alpha,\alpha\rangle\in 2\bz+a \} \ee
be the subsets of points in $\Lambda$ of even and odd
lengths. Then $V^a = V(\Lambda^a)$. It is a well known
fact that if $\Lambda$ is integral, $V(\Lambda)$ is
a vertex super-algebra. It will be a $\bz_2\times\bz_2$
graded vertex algebra iff there are exactly two
(and not more) basisvectors among the $e^i$, say
$e^p$ and $e^q$, for which 
$\langle e^p,e^q\rangle\in\bz+\frac{1}{2}$.
The $\bz_2\times\bz_2$ grading is then defined as follows.
Let $\Lambda^+$ be the lattice that is the $\bz$-span
of the $e^i$, but with $e^p$ and $e^q$ replaced by
$2e^p,2e^q$. (This sublattice will be integral).
Then define the following four vectors
\be v_{00}=0, v_{01}=e^p, v_{10}=e^q, v_{11}=e^p+e^q \ee
Then $\Lambda$ can be written as the union of four
translates of $\Lambda^+$, and this defines the gradation
\be V_{\gamma} = V(\Lambda^+ + v_{\gamma}) \ee
The algebra contains the two special vectors, the identity
${\bf 1}=|0\rangle$, and the Virasoro element.
This element can be written as
\be \omega = \frac{1}{2} \sum_{i,j} g_{ij} \alpha^i_{-1}
   \alpha^j_{-1} |0\rangle \ee
This definition gives $\alpha^i(z)$ conformal weight $1$, 
rank $(V)=c=n$, and
\be L_0 |\beta\rangle = \frac{\langle\beta,\beta\rangle}{2}
  |\beta\rangle \ee
I will call $\omega$ as described here the standard
Virasoro element. Using the tensor product structure of $V$, it is
not hard to derive the following formula for the partition function
\be \mbox{Tr}_V q^{L_0-c/24} = \frac{
  \sum_{\alpha\in\Lambda} q^{\langle\alpha,\alpha\rangle/2}}
  {q^{c/24} \prod_{k=1}^{\infty} (1-q^k)^c } =
  \frac{\theta_{\Lambda}(q)}{\eta(q)^n} \ee
Note that this function is well defined only if
the lattice is positive definite.
The choice of $\omega(z)$ as described above is not the only
one that is possible. For example, for every 
$\alpha\in\Lambda$, the operator 
$\omega'(z)=\omega(z)+\partial^2\alpha(z)$
also satisfies the definition. This possibility
is used in the construction of the ghost-sector of
the superstring. With this definition,
the $\alpha^i(z)$ still have conformal weight one, but
\be L'_0 |\beta\rangle = \frac{\langle\beta-2\alpha,\beta\rangle}{2}
  |\beta\rangle \ee

\subsection{Tensor product algebras.}

Suppose $V(\Lambda_1), V(\Lambda_2)$ are both vertex
algebras. One of them might be $\bz_2\times\bz_2$-graded.
Then their tensor product $V(\Lambda_1)\otimes V(\Lambda_2)$
can be given the structure of a vertex algebra, by saying
\be V(\Lambda_1)\otimes V(\Lambda_2) =
   V(\Lambda_1 \otimes \Lambda_2) \ee
Here by $\Lambda_1 \otimes \Lambda_2$ I mean the orthogonal
product of the two lattices. The effect is that the
bosonic fields used in the construction of the two
factors mutually commute. However, care must be taken for
the operators $\hat{e}_{\alpha}$. The correct prescription
is that if $\alpha\in\Lambda_1$ and $\beta\in\Lambda_2$,
then one should have
\be \hat{e}_{\alpha}\hat{e}_{\beta}=(-1)^{
    \langle\alpha,\alpha\rangle \langle\beta,\beta\rangle}
    \hat{e}_{\beta}\hat{e}_{\alpha} \ee

\section{Toroidal compactified superstring.}
\setcounter{equation}{0}

In this section I give a description of the
left moving algebra of the superstring, compactified
on the unique Lorentzian lattice in ten dimensions
$E_{10}$, in the RNS formulation, in terms of vertex algebras.
I will choose a basis for the coordinates of the string
in which $E_{10}$ splits into an orthogonal product
$E_8 \otimes E_{1,1}$, and then describe the
coordinates compactified on $E_8$ (the transverse part)
and those on $E_{1,1}$ (the lightcone) separately.  

\subsection{The transverse part.}

I will restrict my attention to the left movers.
The bosonic fields, compactified on $E_8$, are 
described by a vertex-algebra $V(E_8)$. I will call
the bosonic fields used in the construction
$\partial X^i(z)$ where $1\leq i\leq 8$.
They satisfy the OPE's
\be \partial X^i(z)\partial X^j(w) =
  \frac{\delta^{i,j}}{(z-w)^2}+\cdots \ee
The fermionic fields are described by a $\bz_2\times\bz_2$-graded vertex
algebra $V(D_4^*)$ (before GSO-projection) \cite{pict,spinor,related}.
Here $D_4^*$ is the weight-lattice of the Lie-algebra $D_4$,
the dual of the root lattice. Let me relate
this compact definition to the concepts from string
theory. First I will define the $\bz_2\times\bz_2$ gradation
on $V(D_4^*)$. A basis for the lattice $D_4$ is given by
\be e^1=\left[ \begin{array}{c} 1\\-1\\0\\0 \end{array} \right],
    e^2=\left[ \begin{array}{c} 0\\1\\-1\\0 \end{array} \right],
    e^3=\left[ \begin{array}{c} 0\\0\\1\\-1 \end{array} \right],
    e^4=\left[ \begin{array}{c} 0\\0\\1\\ 1 \end{array} \right] \ee
(in terms of the basis $f^i$, with $\langle f^i,f^j\rangle=\delta^{i,j}$)
and the Gramm-matrix then becomes (giving actually the
Cartan matrix for the Lie-algebra $D_4$)
\be \langle e^i,e^j\rangle = \left( \begin{array}{cccc}
   2&-1&0&0 \\ -1&2&-1&-1 \\ 0&-1&2&0 \\
   0&-1&0&2 \end{array} \right) \ee
The dual of $D_4$ can be written as a union of four
translates of $D_4$ \cite{sloane}
\be D_4^* = \bigcup_{\gamma\in\bz_2\times\bz_2} 
  (D_4 + \Delta_{\gamma}) \ee
where the translation vectors are given by
\be \Delta_{00}=\left[ \begin{array}{c} 0\\0\\0\\0 \end{array} \right],
\Delta_{01}=\left[ \begin{array}{c} 1\\0\\0\\0 \end{array} \right],
\Delta_{10}=\left[ \begin{array}{c} \frac{1}{2}\\ \frac{1}{2}\\
       \frac{1}{2}\\ \frac{1}{2} \end{array} \right],
\Delta_{11}=\left[ \begin{array}{c} -\frac{1}{2}\\ \frac{1}{2}\\
       \frac{1}{2}\\ \frac{1}{2} \end{array} \right] \ee
The translation vectors are labeled by elements of the
group $\bz_2\times\bz_2$ because $\Delta_{\gamma_1}+\Delta_{\gamma_2}
=\Delta_{\gamma_1+\gamma_2}$ modulo elements of $D_4$.
(They are actually representants of the quotient $D_4^*/D_4$).
This defines the $\bz_2\times\bz_2$-gradation of $V(D_4^*)$
\be V(D_4^*) = \bigoplus_{\gamma\in\bz_2\times\bz_2}
  \hat{\Lambda}_{D_4+\Delta_{\gamma}} \otimes H^4 
  = \bigoplus_{\gamma\in\bz_2\times\bz_2} V_{\gamma} \ee
The subspace $V_{00}\oplus V_{01}$
is the vertex super-algebra $V(\bz^4)$. It is a well known result
of boson-fermion correspondence theorems \cite{fons} that
this algebra is isomorphic to the algebra constructed
from eight fermionic fields with NS-boundary conditions.
The correspondence is as follows. The vectors
$\beta^i = f^i$, with $1\leq i\leq 4$ form a basis for
this lattice $\bz^4$. Then the vertex operators defined by
\be \psi^i(z) = \no{ e^{\beta^i(z)} }, \mbox{\hspace{1cm}}
    \bar{\psi}^i(z) = \no{ e^{-\beta^i(z)} } \ee
are complex fermions. They satisfy OPE's
\be \psi^i(z)\bar{\psi}^j(w) =
  \frac{\delta^{i,j}}{(z-w)} + \cdots \ee
For later use, I will define real fermions
in terms of these
\be b^{2i-1}(z) = \frac{\psi^i(z)+\bar{\psi}^i(z)}{\sqrt{2}},
  \mbox{\hspace{1cm}}
    b^{2i}(z) = \frac{\psi^i(z)-\bar{\psi}^i(z)}{i\sqrt{2}} \ee
which satisfy
\be b^i(z)b^j(w) = 
  \frac{\delta^{i,j}}{(z-w)} + \cdots \ee
So the space $V_{00}\oplus V_{01}$ can be identified
as the NS-sector of the string theory. Now one can check 
that the operators
$\psi^i,\bar{\psi}^i$ have an expansion in half
integral powers of $z$ when acting on $V_{10}$ or
$V_{11}$. So $V_{10} \oplus V_{11}$ can be identified
as the R-sector. The operators $\hat{e}_{\Delta_{10}}$ and
$\hat{e}_{\Delta_{11}}$ are examples of spectral
flow operators, they are invertible operators
that map NS-states to R-states and vice versa.
The tensor product of the fermionic
and bosonic algebras are a $\bz_2\times\bz_2$-graded
vertex algebra, and I define
\be V^T_{\gamma} = V(E_8) \otimes V(D_4+\Delta_{\gamma}) \ee
Note that this division into a NS- and a R-sector
depends upon the choice which element $V_{\gamma}$
with $\gamma\in\{01,10,11\}$ is added to
$V_{00}$ to form a vertex super-algebra. The
fermions in this algebra still have NS-boundary
conditions when acting on $V_{00}\oplus V_{\gamma}$,
and R-boundary conditions when acting on the
remaining two $V's$. In the algebra $V(D_4^*)$,
this is of no consequence, because the three
possible choices are related by triality \cite{spinor},
but in general it does matter.
%

\subsection{Partition functions.}

It is easy to obtain the partition functions
for the algebras constructed in the previous section.
For the bosonic part one finds
\be \mbox{Tr}_{V(E_8)} q^{L_0-1/3} = \frac{\theta_{E_8}(q)}{\eta^8(q)}
  = \frac{E_4(q)}{\eta^8(q)} = q^{-1/3}(1+248q+\cdots) \ee
where $E_4(q)$ is the normalized Eisenstein series of weight $4$.
As for the fermionic part, these are given by
\be P_{\gamma}(q) = \mbox{Tr}_{V(D_4+\Delta_{\gamma})}
  q^{L_0-1/6} = \frac{\theta_{D_4+\Delta_{\gamma}}(q)}{\eta^4(q)} \ee
From triality it follows that the $P_{\gamma}$ for
$\gamma \neq 00$ are all equal. Now from the fermionic
formulation of the superstring, or by using some modular
identities, one finds the familiar result \cite{gsw}
\be P_{\gamma}(q) = 8q^{1/3}\prod_{k=1}^{\infty}(1+q^k)^8,
  \mbox{\hspace{1cm}for $\gamma\neq 00$} \ee
These partition functions can be refined by introducing
a second gradation in the algebra $V(D_4^*)$, leading
to a chiral version of the elliptic genus. This gradation
will be defined by a $U(1)$ subalgebra of the $D_4$
algebra that acts on $V(D_4^*)$. It requires the choice
of a specific weight one vertex-operator. There are
three preferred choices that can be made, namely the
bosonic fields $J^{\gamma}(z)$ associated with
the lattice points $2\Delta_{\gamma} \in D_4$,
with $\gamma\neq 00$. Their zero modes act like
\be J_0^{\gamma} | \beta \rangle = 2\langle\Delta_{\gamma},
  \beta\rangle |\beta\rangle \ee
and they commute with the Heisenberg-algebra.
Among these, $J_0^{11}$ is special, because $J^{11}(z)$
is the $U(1)$ current that combines with the $N=1$ algebra,
described in section $4$, to give the standard $N=2$
algebra in the transverse theory. This $U(1)$ current
remains well defined in more general compactifications
with space-time supersymmetry \cite{sptm}.
Furthermore, $(-1)^{J_0^{11}}$ defines 
(part of) the GSO-projection. It acts as
$+1$ on $V^T_{00},V^T_{11}$, and as $-1$ on
$V^T_{01},V^T_{10}$. So let me write simply
$J_0=J_0^{11}$. Now one can define the functions
\be P_{\gamma}(q,y) = \mbox{Tr}_{V(D_4+\Delta_{\gamma})}
  q^{L_0-1/6} y^{J_0} = \frac{\sum_{\alpha\in D_4+\Delta_{\gamma}}
  q^{\langle\alpha,\alpha\rangle/2} y^{2\langle\Delta_{11},\alpha\rangle}}
  {\eta^4(q)} \ee
which reduces to $P_{\gamma}(q)$ for $y=1$. 
Again, using the fermionic formulation, these functions
can be expressed in a product form. The result is
\be P_{\gamma}(q,y) = \left\{ \begin{array}{ll}
  \theta_1^+(q,y) & \mbox{if $\gamma = 00$} \\
  \theta_1^-(q,y) & \mbox{if $\gamma = 11$} \\
  \theta_2(q,y)   & \mbox{otherwise} \end{array} \right. \ee
where
\[ \theta^{\pm}_1 =q^{-1/6}\prod_{n=0}^{\infty}
(1+q^{n+\frac{1}{2}}y^{-2})(1+q^{n+\frac{1}{2}})^6
(1+q^{n+\frac{1}{2}}y^2) \]
\be \pm q^{-1/6} \prod_{n=0}^{\infty}
(1-q^{n+\frac{1}{2}}y^{-2})(1-q^{n+\frac{1}{2}})^6
(1-q^{n+\frac{1}{2}}y^2) \ee
\be \theta_2 = q^{1/3}(4y^{-1}+4y)\prod_{n=1}^{\infty}
(1+q^ny^{-2})(1+q^n)^6(1+q^ny^2) \ee
These functions have the following series
expansions
\be \theta_1^+(q,y)=q^{-1/6}(1+(6y^{-2}+16+6y^2)q+\cdots) \ee
\be \theta_1^-(q,y)=q^{1/3}((y^{-2}+6+y^2)+(16y^{-2}+32+16y^2)q
  +\cdots) \ee
\be \theta_2(q,y)=q^{1/3}((4y^{-1}+4)+(4y^{-3}+28y^{-1}+28y+4y^3)q
  +\cdots) \ee
By using the spectral flow operator $\hat{e}_{\Delta_{11}}$
one can find the following identity
\[ P_{\gamma+11}(q,y) = \mbox{Tr}_{V(D_4+\Delta_{\gamma})}
  \hat{e}^{-1}_{\Delta_{11}} q^{L_0-1/6} y^{J^{11}_0}
  \hat{e}_{\Delta_{11}} = \]
\be \mbox{Tr}_{V(D_4+\Delta_{\gamma})}
  q^{L_0+\frac{1}{2}J^{11}_0+1/3} y^{J^{11}_0+2} = 
  \sqrt{q}y^2 P_{\gamma}(q,\sqrt{q}y) \ee
In general, applying spectral flow $n$ times to
a vector with $(L_0,J_0)$-eigenvalues $(q,k)$ sends
it to a vector with eigenvalues 
$(q+\frac{nk}{2}+\frac{n^2}{2},k+2n)$ \cite{chiral}
(iff $c=12$). So let $(q,-k)$ be a vector in
$V_{01}$. Then $k$ must be odd, and applying
spectral flow $k$ times sends it to a vector
$(q,k)$ in $V_{10}$. This leads to the identity
\be P_{01}(q,y) = P_{10}(q,y) \label{eqsf2} \ee
These spectral flow arguments of course remain
valid for more general compactifications. 
In the GKM-algebras that will be constructed, I will be interested
only in the partition functions of $V^T_{01}$, which will count
bosons, and $V^T_{11}$, which counts fermions
(with given momentum). So one can define a
'supersymmetric index' analogously to \cite{sidx}
\be I(q,y) = \mbox{Tr}_{V^T_{01}\oplus V^T_{11}}
   (-1)^{J_0} q^{L_0-1/2} y^{J_0} = \frac{E_4(q)}{\eta^8(q)}
   (\theta_1^-(q,y)- \theta_2(q,y) ) \ee
By using (\ref{eqsf2}), this function can also
be written as a trace over the R-sector, making
it look like a chiral version of the elliptic genus
\be I(q,y) = \mbox{Tr}_{V^T_{10}\oplus V^T_{11}}
   (-1)^{J_0} q^{L_0-1/2} y^{J_0} \ee
By using spectral flow it can also be expressed 
as a trace over the NS-sector
\be I(q,y) = \mbox{Tr}_{V^T_{00}\oplus V^T_{01}}
   (-1)^{J_0} q^{L_0+J_0} y^{J_0+2} \ee
The function turns out to have all the defining properties of a
so called weak Jacobi form of index $2$ and weight $0$
\cite{EZ,mumf,me}. The space of such forms is
two dimensional, with basis
\be J_1 = E_4(q)\left(\frac{\phi_{10,1}(q,y)}{\eta^{24}(q)}\right)^2
  = (y^{-2}-4y^{-1}+6-4y+y^2) + O(q) \ee
\be J_2 = \left(\frac{\phi_{12,1}(q,y)}{\eta^{24}(q)}\right)^2
  = (y^{-2}+20y^{-1}+102+20y+y^2) + O(q) \ee
Here $\phi_{10,1}$ and $\phi_{12,1}$ are unique cusp forms of
index $1$ and weights $10$ and $12$ respectively, described in \cite{EZ}.
Comparing a few coefficients shows that $I=J_1$.

\subsection{The lightcone and ghosts.}

In this section I describe the lightcone coordinates
of the superstring plus the ghost-sector. I describe them
together, because only together do they form proper vertex algebras.
I claim that this algebra can be described as (a subalgebra of)
the tensorproduct of four vertex algebras, namely
$V(E_{1,1}) \otimes V_1(\bz) \otimes V_2(\bz) \otimes V(L)$.
I will now describe where these factors come from.

\vspace{1\baselineskip} \noindent
The vertex algebra $V(E_{1,1})$ corresponds to the
bosonic coordinates of the superstring, compactified on
$E_{1,1}$. This lattice has Gramm-matrix
\be \left( \begin{array}{cc} 0&1\\1&0 \end{array} \right) \ee
I will call the bosonic fields used in the construction
$\partial X^0(z),\partial X^9(z)$, and they satisfy
\be \partial X^i(z) \partial X^j(w) =
  \frac{\eta^{ij}}{(z-w)^2} + \cdots \ee
where $\eta$ is diagonal, $-\eta^{00}=\eta^{99}=1$.

\vspace{1\baselineskip} \noindent
The two factors $V_{1,2}(\bz)$ describe the $b,c$-ghostsystem
and the $\eta,\xi$ system respectively, which is part of the
$\beta,\gamma$-ghostsystem \cite{pict}. Here the lattice $\bz$ is
the lattice with basis $e^1=f^1$, and $\langle f^1,f^1\rangle = 1$.
Now let $\sigma=f^1$ be in the lattice coming from $V_1(\bz)$
and $\chi=f^1$ in the lattice coming from $V_2(\bz)$. Then
\be c(z) = \no { e^{\sigma(z)} }, \mbox{\hspace{1cm}}
    b(z) = \no { e^{-\sigma(z)} } \ee
\be \xi(z) = \no { e^{\chi(z)} }, \mbox{\hspace{1cm}}
    \eta(z) = \no { e^{-\chi(z)} } \ee
The $b,c$-system satisfies the OPE's (and similar for $\xi,\eta$)
\be c(z)b(w) = \frac{1}{(z-w)} + \cdots \ee
The factor $V(L)$ is a $\bz_2\times\bz_2$ graded vertex
algebra. The lattice $L$, which has signature $(1,1)$,
can be written as the union of four translates of an
even sublattice $L^+$. A basis for $L^+$ is given by
\be e^1=\left[ \begin{array}{c} 1\\-1 \end{array} \right],
    e^2=\left[ \begin{array}{c} 1\\1 \end{array} \right] \ee
in terms of the basis $f^1,f^2$, where
$\langle f^i,f^j\rangle=\mbox{diag}(-1,1)$.
Then $L$ is described by
\be L = \bigcup_{\gamma\in\bz_2\times\bz_2}
  (L^+ + \delta_{\gamma}) \ee
where the translation vectors are given by
\be \delta_{00}=\left[ \begin{array}{c} 0\\0 \end{array} \right],
\delta_{01}=\left[ \begin{array}{c} 1\\0 \end{array} \right],
\delta_{10}=\left[ \begin{array}{c} \frac{1}{2}\\ \frac{1}{2}
       \end{array} \right],
\delta_{11}=\left[ \begin{array}{c} -\frac{1}{2}\\ \frac{1}{2}
       \end{array} \right] \ee
This defines the $\bz_2\times\bz_2$-gradation of $V(L)$,
just as in the case of $V(D_4^*)$
\be V(L) = \bigoplus_{\gamma\in\bz_2\times\bz_2}
     V(L^+ + \delta_{\gamma}) =
      \bigoplus_{\gamma\in\bz_2\times\bz_2} V_{\gamma} \ee
Clearly, the lattice $L$ contains the points
$\phi=f^1$, and $\rho=f^2$. I can now identify
\be b^+(z) = \no { e^{\rho(z)} }, \mbox{\hspace{1cm}}
    b^-(z) = \no { e^{-\rho(z)}} \ee
\be \gamma(z) = \no {e^{\phi(z)}\eta(z) }, \mbox{\hspace{1cm}}
    \beta(z) = \no {e^{-\phi(z)}\partial\xi(z) } \ee
The fermionic fields $b^{\pm}(z)$ satisfy
\be b^+(z) b^-(w) = \frac{1}{(z-w)} + \cdots \ee
They are the lightcone fermions of the matter-system.
The bosonic fields $\beta(z),\gamma(z)$ satisfy
\be \gamma(z) \beta(w) = \frac{1}{(z-w)} + \cdots \ee
Note that $\no{ \gamma(z)\beta(z) } = -\phi(z)$ and
$\no{ c(z)b(z) } = \sigma(z)$. For later use I define
\be b^{0}(z) = \frac{b^+(z)-b^-(z)}{\sqrt{2}},
  \mbox{\hspace{1cm}}
    b^{9}(z) = \frac{b^+(z)+b^-(z)}{\sqrt{2}} \ee
In $V(L)$ there is no ambiguity about what to call the
NS-sector, and what the R-sector, because only
$V_{00}$ and $V_{01}$ contain states without 
ghost-excitations, so they must be the NS-sector.
Also, the entire lightcone plus ghosts algebra
is again a $\bz_2\times\bz_2$ graded vertex algebra,
and I define
\be V^L_{\gamma} = V(E_{1,1}) \otimes
  V_1(\bz) \otimes V_2(\bz) \otimes V_{\gamma} \ee 

\subsection{Fitting it together.}

The NS-sector of the total left moving algebra
is of course the product of the NS-sectors of
transverse part and the lightcone plus ghost part,
and similar for the R-sector. Thus
\be V_{NS} = (V^L_{00} \oplus V^L_{01})\otimes
      (V^T_{00} \oplus V^T_{01}) \ee
\be V_{R} = (V^L_{10} \oplus V^L_{11})\otimes
      (V^T_{10} \oplus V^T_{11}) \ee
The reader can check that this
space again has the structure of a
$\bz_2\times\bz_2$ graded vertex algebra,
with a gradation defined by
\be V_{00} = V^L_{00}\otimes V^T_{00} \oplus
             V^L_{01}\otimes V^T_{01} \ee
\be V_{01} = V^L_{00}\otimes V^T_{01} \oplus
             V^L_{01}\otimes V^T_{00} \ee
\be V_{10} = V^L_{10}\otimes V^T_{10} \oplus
             V^L_{11}\otimes V^T_{11} \ee
\be V_{11} = V^L_{10}\otimes V^T_{11} \oplus
             V^L_{11}\otimes V^T_{10} \ee
I will call this algebra the superstring vertex-algebra.
Then the subspace $V_{00}\oplus V_{10}$ is
a vertex super-algebra, and this is exactly the
space that is projected out by the GSO-projection \cite{BRST}.

\section{The BRST-construction.}
\setcounter{equation}{0}

In order to define the physical states of the string,
one needs to define the BRST-operator $Q$. This operator
is defined in terms of the $N=1$ supersymmetry algebra
that is present in the superstring algebra \cite{gsw,BRST}.
The $N=1$ algebra is generated by two fields
$T(z), G(z)$, that must satisfy the following OPE's
\be T(z)T(w) = \frac{c}            {(z-w)^4}
             + \frac{2T(w)}        {(z-w)^2}
             + \frac{\partial T(w)}{(z-w)} + \cdots \ee
\be T(z)G(w) = \frac{3G(w)/2}      {(z-w)^2}
             + \frac{\partial G(w)}{(z-w)} + \cdots \ee
\be G(z)G(w) = \frac{2c/3}         {(z-w)^3}
             + \frac{2T(w)}        {(z-w)} + \cdots \ee
where $c$ is some constant. This structure is
present in both the matter sector and the ghost sector.
In the matter sector it is given by
\be T_m(z) =
  \frac{1}{2} \eta_{ij} \no { (\partial X^i) (\partial X^j) 
  + (\partial b^i) b^j } \ee
\be G_m(z) =
  \eta_{ij} (\partial X^i) b^j \ee
with $c=15$. $T_m(z)$ is just the operator $\omega(z)$
corresponding to the standard Virasoro element in this part
of the algebra. In the ghostsector one has
\be T_g(z) = - \no{ (\partial b)c }
             -2\no{ b \partial c  }
   -\frac{3}{2}\no{ \beta \partial\gamma }
   -\frac{1}{2}\no{ (\partial\beta) \gamma } \ee
\be G_g(z) = -2 b\gamma + c\partial\beta
   + \frac{3}{2} (\partial c)\beta \ee
with $c=-15$. In this case $T_g(z)$ is not equal
to $\omega(z)$, but is actually given by
\be T_g(z) = \omega(z) + \no { \partial (
  \frac{3}{2} cb - \gamma\beta ) } \ee
This gives the fields $b,c,\beta,\gamma$ conformal
weights $2,-1,\frac{3}{2},-\frac{1}{2}$ respectively.
The operators $T_m$ and $T_g$ are in $V_{00}$,
while the operator $G_m$ and $G_g$ are in $V_{01}$.
The BRST-current $Q(z)$ is defined as
\be Q(z) = \no { (T_m+\frac{1}{2}T_g)c
    + (G_m+\frac{1}{2}G_g)\gamma } \ee
This fermionic field has conformal weight $1$.
Its zero mode $Q_0$, which I will write as $Q$,
satisfies \cite{BRST}
\be Q^2 = 0 \ee
Since $Q(z)$ is in $V_{00}$, it maps every
$V_{\gamma}$ to itself. To define the cohomology
problem, I have to define some subspaces of the
superstring algebra. First the factor $V_2(\bz)\otimes V(L)$
in the construction of this algebra is restricted
to a subalgebra $W$, which is graded by the 
ghostpicture (number) \cite{pict}. This subalgebra
can be defined as follows. In the product lattice
$L\otimes\bz$ one can define a set of points
given by $v_k = 2k\delta_{10}$ with $k\in\bz/2$
(These vectors do not have a component in the $\bz$
lattice direction). These points define a set
of vectors $|v_k\rangle$
in the algebra $V_2(\bz)\otimes V(L)$.
From every $|v_k\rangle$, I can construct a space
$W_k$, by letting the component operators
of the fields $b^{\pm}(z),\beta(z),\gamma(z)$
act on it in all possible ways. The number $k$
will be called the ghostpicture.  I now define
\be W = \bigoplus_{k\in\bz/2} W_k \ee
One can easily see that the subalgebra $W$ is again
a $\bz_2\times\bz_2$ graded vertex algebra, and that the
ghostpicture defines a gradation of it.
This then defines a subalgebra of the superstring
algebra (by using $W$ instead of $V_2(\bz)\otimes V(L)$
in its construction), which is also graded by
ghostpicture number. Let me call the graded pieces of 
this subalgebra $V^s_{\gamma,k}$. Note that
$k\in\bz+\Theta(\gamma)$ where
\be \Theta(\gamma)=\left\{ \begin{array}{ll}
  0 & \mbox{if $\gamma=00,01$ (NS sector)} \\
  \frac{1}{2} & \mbox{if $\gamma=10,11$ (R-sector)}
  \end{array} \right. \ee
Since $Q$ has ghostpicture number $0$, it leaves all
$V^s_{\gamma,k}$ invariant.
Next, I need another gradation, by ghostnumber
(not to be confused with ghostpicture).
This gradation is defined by the zero-mode of the
ghostnumber current defined by \cite{BRST}
\be J(z) = \no{ cb - \gamma\beta } =\sigma(z)+\phi(z) \ee
It gives $c,\gamma$ ghostnumber $1$ and
$b,\beta$ ghostnumber $-1$. Consequently $Q$ has 
ghostnumber $1$. In general, the ghostnumber will
be in $\bz+\Theta(\gamma)$ for $V^s_{\gamma,k}$.
The final gradation that I will
use is the gradation by the lightcone-lattice
$E_{1,1}$. It will in the end define a gradation
on the GKM-algebras that I construct. Let me write
\be V^s_{\gamma,k} = \bigoplus_{\alpha\in E_{1,1},
  n\in\bz+\Theta(\gamma)} V^s_{\gamma,k}(\alpha,n) \ee
where $n$ is the ghostnumber. I am now ready to
define the cohomology problem. What I'm interested in is
the relative cohomology of Lian and Zuckerman \cite{BRST}.
This is $Q$-cohomology in a subspace of 
$V^s_{\gamma,k}$, defined by
\be \ker L_0 \cap \ker b_0 \ee
in the NS-sector ($\gamma=00,01$), and
\be \ker L_0 \cap \ker b_0 \cap \ker G_0 \cap \ker \beta_0 \ee
in the R-sector ($\gamma=10,11$). Here $L_0,b_0,G_0,\beta_0$
are the zero-modes of the fields $T_g(z)+T_m(z),b(z),
G_g(z)+G_m(z),\beta(z)$ respectively. Note that $G_0$ and
$\beta_0$ only exist as operators in the R-sector. Let
$C_{\gamma,k}$ be this subspace. The cohomology is
\be H_{\gamma,k}(\alpha,n) =
  \ker_Q C_{\gamma,k}(\alpha,n) / \mbox{im}_Q
     C_{\gamma,k}(\alpha,n-1)  \ee
(This is to be contrasted with the absolute cohomology 
$H^{\mbox{\scriptsize abs}}$, which is the $Q$-cohomology of $V^s_{\gamma,k}$).
The results of this calculation are well known \cite{pict,BRST}.
I will state the results. There are two cases to consider,
$\alpha=0$ and $\alpha\neq 0$. For $\alpha\neq 0$, one has
\be H_{\gamma,k}(\alpha,n) = 0, 
   \mbox{\hspace{1cm}iff $n-k\neq 1$} \ee
\be \dim H_{\gamma,k}(\alpha,k+1) = 
    \dim (V^T_{\gamma+01})_{-\frac{\langle\alpha,\alpha\rangle}{2}
    +\frac{1}{2}} \ee
where $(V^T_{\gamma})_n$ is the subspace of $V^T_{\gamma}$
of conformal dimension $n$. If $\alpha=0$, one finds
\be H_{\gamma,k}(0,n) = 0,
   \mbox{\hspace{1cm}iff $n-k\neq 0,1,2$} \ee
\be \dim H_{\gamma,k}(0,k) = \dim H_{\gamma,k}(0,k+2) 
 = \left\{ \begin{array}{ll} 1 & \mbox{NS-sector} \\
            0 & \mbox{R-sector} \end{array} \right. \ee
\be \dim H_{\gamma,k}(0,k+1) = \left\{ \begin{array}{ll}
   10 & \gamma=00 \\ 0 & \gamma=01 \\
   16 & \gamma=10 \\ 16 & \gamma=11 \end{array} \right. \ee
After the GSO-projection, which projects out $V_{00}\oplus V_{10}$,
the physical states of the superstring are given by
\be H = \left( \bigoplus_{k\in\bz,\alpha\in E_{1,1}}
  H_{00,k}(\alpha,k+1) \right) \oplus \left(
  \bigoplus_{k\in\bz+\frac{1}{2},\alpha\in E_{1,1}}
  H_{10,k}(\alpha,k+1) \right) \ee
modulo the ghostpicture equivalence \cite{pict},
to be described in the next section. Note that
representatives of $H_{00,k}(\alpha,k+1)$ are
fermionic states (in the vertex algebra) and
representatives of $H_{10,k}(\alpha,k+1)$ are bosonic.

\section{Gerstenhaber algebra.}
\setcounter{equation}{0}

In this section I describe the Gerstenhaber-algebra
which is present in the absolute superstring cohomology.
Let me recall the definition of such an algebra \cite{LiZu}.
It is a $\bz_2 \times \bz$-graded vector space $V$
\be V = \bigoplus_{a\in\bz_2,m\in\bz} V^a_m \ee
equipped with two multiplication operations,
the dot product
\be \cdot: V^a_m \times V^b_n \rightarrow
   V^{a+b}_{m+n} \ee
and the bracket
\be \{,\}: V^a_m \times V^b_n \rightarrow
  V^{a+b-1}_{m+n-1} \ee
Let $u\in V^a_m, v\in V^b_n, w\in V^c_p$. The
dot product is associative commutative
\be u\cdot v = (-1)^{ab} v\cdot u \ee
\be (u\cdot v)\cdot w = u\cdot (v\cdot w) \ee
The bracket satisfies
\be \{u,v\}=-(-1)^{(a-1)(b-1)} \{v,u\} \ee
\[ (-1)^{(a-1)(c-1)}\{u,\{v,w\}\}+
   (-1)^{(c-1)(b-1)}\{w,\{u,v\}\} \]
\be +(-1)^{(b-1)(a-1)}\{v,\{w,u\}\}=0  \ee
(It is like a Lie-bracket with the definitions
of fermions and bosons swapped). Finally there
is a relation between the bracket and the dot product
\be \{u,v\cdot w\} = \{u,v\}\cdot w+
   (-1)^{(a-1)b} v\cdot\{u,w\} \ee

\subsection{The multiplications.}

The multiplications that define the Gerstenhaber algebra
are just Lian and Zuckerman's dot product and bracket,
which they use for bosonic strings \cite{LiZu}.
Let me recall their definitions. The bracket is given by
\be \{,\}: V^s_{\gamma_1,k}(\alpha,m) \times
  V^s_{\gamma_2,l}(\beta,n) \rightarrow
  V^s_{\gamma_1+\gamma_2,k+l}(\alpha+\beta,m+n-1) \ee
\be \{ u,v \} = (-1)^a \mbox{Res}_w\mbox{Res}_{z-w}
  (b(z-w)u)(w)v \ee
with $u\in V^a$. The dot product is
\be \cdot:  V^s_{\gamma_1,k}(\alpha,m) \times
  V^s_{\gamma_2,l}(\beta,n) \rightarrow
  V^s_{\gamma_1+\gamma_2,k+l}(\alpha+\beta,m+n) \ee
\be u\cdot v=\mbox{Res}_z\frac{u(z)v}{z} \ee
Both products are well defined iff 
$\Delta(\gamma_1,\gamma_2)=0$. As was proven by
Lian and Zuckerman, these products define 
products on the $Q$-cohomology, satisfying
all definitions of a Gerstenhaber algebra. The $\bz$
gradation in the definition of a Gerstenhaber algebra
is given by (ghostnumber minus ghostpicture). The bracket
has the nice property that it closes on the
relative cohomology. Furthermore, if $u$ and $v$
both are in $\ker b_0$ there is the following
relation between the dot product and the bracket
\be \{u,v\} = b_0 (u\cdot v) \label{dotbrak} \ee
The dot product does not close on the relative
cohomology. The last ingredient that I need
to define the Lie algebra on the physical states
is the picture change operation. This is an
isomorphism $\phi$ on the absolute cohomology
relating different ghostpictures
\be \phi: H^{\mbox{\scriptsize abs}}_{\gamma,k}(\alpha,m)
  \rightarrow
  H^{\mbox{\scriptsize abs}}_{\gamma,k+1}(\alpha,m+1) \ee
For the construction of this isomorphism I refer to the
literature \cite{pict}, but note that it is compatible with
the dot and bracket product
\be \phi( \{ u,v\} ) = \{ \phi(u), v \} \bmod Q \ee
\be \phi( u\cdot v ) = \phi(u)\cdot v \bmod Q \ee
This thus defines the products on the absolute cohomology
modulo ghost-picture equivalence, and in particular it
defines a Lie-algebra on the physical states $P=H/\phi$.
(Note that representatives of cohomology classes in the NS-sector
where fermionic in the vertex algebra. They are now
bosons in the Lie algebra, as they should). The Lie-algebra
is graded by the lattice $E_{1,1}$
\be P = \bigoplus_{\alpha\in E_{1,1}} P_{\alpha} \ee

\subsection{The massless left-moving algebra.}

In this section I give a description of the Lie-algebra
on the left-moving massless modes. As is well known \cite{gsw},
such states can be described by their momentum $p$ with
$p\cdot p=0$, and by a polarization vector $k$ in the NS-sector,
or by a spinor $u$ in the R-sector. This corresponds
to the fact that the massless states in the NS-sector transform
in the vector representation of the $SO(9,1)$ symmetry algebra
that acts on the superstring algebra. The massless states in the
R-sector transform in the spinor representation.
I will denote these states by $(k,p)$ and $(u,p)$ respectively.
In ghostpicture $0$, the state $(k,p)$ is represented by
\be \left( c(z)k\cdot\partial X(z)+c(z)(p\cdot b(z))(k\cdot b(z))
   +\gamma(z) k\cdot b(z) \right) e^{p\cdot X} \ee
The polarization vector must satisfy $p\cdot k=0$.
The same state is represented in ghostpicture $-1$ by
\be c(z)(k\cdot b(z))e^{p\cdot X}e^{-\phi} \ee
The commutator
\be [Q,e^{p\cdot X}] = \left( \frac{(p\cdot p)\partial c(z)}{2}
   + c(z)p\cdot \partial X(z)+\gamma(z) p\cdot b(z)\right) 
   e^{p\cdot X} \ee
shows that $k \sim k+\alpha p$ with $\alpha\in {\bf R}$ 
in $Q$-cohomology. The state $(u,p)$ is represented
in ghostpicture $-\frac{1}{2}$ by
\be c(z)u^{\alpha}S_{\alpha}(z)e^{p\cdot\phi}e^{-\phi/2} \ee
The spinor $u$ must satisfy the Dirac-equation $\hat{p} u=0$.
Here $\hat{p}=p_{\mu} \gamma^{\mu}$, and the $\gamma^{\mu}$
are Dirac matrices satisfying $\{ \gamma^{\mu},
\gamma^{\nu} \}=2\eta^{\mu\nu}$. I can now calculate
the bracket on the massless states by plugging these
representatives into the definition of the bracket.
I will not go into detail, but only state the results.
Note that the bracket of two states with momenta $p_1,p_2$
only gives a massless state iff $(p_1+p_2)^2=0$, i.e.
$p_1 \cdot p_2=0$. Under this condition one finds
\be [ (k_1,p_1), (k_2,p_2)] =
  ( (k_1\cdot p_2)k_2-(k_2\cdot p_1)k_1+
    (k_1\cdot k_2)p_1, p_1+p_2) \ee
At first sight, this bracket appears not to be
antisymmetric. However it is antisymmetric modulo states
of the form $(p,p)$, which are $Q$-exact.
\be [ (k_1,p_1) , (u_2,p_2) ] =
  ( (k_1\cdot p_2)u_2 + \frac{1}{2}\hat{p}_1\hat{k}_1u_2,
  p_1+p_2) \ee
\be [ (u_1,p_1), (u_2, p_2) ] =
      (u_1\gamma u_2, p_1+p_2) \ee
One can check that the bracket is indeed a Lie-bracket,
(as far as it closes on massless states),
where the $(k,p)$ are bosonic, and the $(u,p)$ are fermionic.

\section{On GKM-algebras.}
\setcounter{equation}{0}

\subsection{Chiral algebras.} \label{sch}

In this section, I consider some possibilities to
construct Lie-algebras. The basic example 
is just the Lie-algebra on the physical states $P$
(the relative cohomology modulo ghostpicture equivalence),
of the chiral (left-moving) string, as described in the
previous section. This Lie-algebra is graded by the
lattice $E_{1,1}$ (it is actually graded by the whole
compactification lattice $E_{10}$) and it is the super
equivalent of the construction of Borcherds \cite{moon}.
It is therefore a GKM-algebra. Note that the
superdimensions of the homogeneous spaces $P_{\alpha}$
vanish unless $\alpha=0$
\be \dim P^0_{\alpha} - \dim P^1_{\alpha} =
  \left\{ \begin{array}{ll} 0 & \alpha\neq 0 \\
  -6 & \alpha=0 \end{array} \right. \ee
Other $E_{1,1}$-graded GKM-algebras can be made by
replacing the transverse algebra by a more general
algebra. This algebra must be a 
$\bz_2\times\bz_2$-graded vertex algebra in order
to make the total GSO-projected algebra a
vertex superalgebra. The most interesting case
is of course when the resulting theory has spacetime
supersymmetry. As is well known \cite{sptm},
this requires a transverse algebra with $N=2$
supersymmetry. This $N=2$ algebra must have central
charge $c=12$ (or $\hat{c}=4$), to have $Q^2=0$.
This replacement leads to a family of $E_{1,1}$-graded
GKM-algebras, whose (graded) dimensions can be expressed
in terms of the supersymmetric index $I(q,y)$ of the
transverse algebra (for $\alpha\neq 0$)
\be \dim P^0_{\alpha} - \dim P^1_{\alpha} =
  -\mbox{Res}_q q^{\frac{\langle\alpha,\alpha\rangle}{2}-1}
  I(q,y=1) \ee
By virtue of supersymmetry, these dimensions are
zero, unless $\langle\alpha,\alpha\rangle=0$.
Note that the $y$-dependence of supersymmetric index
does not play a role in this Lie-algebra, there is
not a gradation of the GKM-algebra corresponding
to the $J_0$ gradation of the transverse algebra.

\subsection{BPS-algebras.}

In this section I consider the algebra structure
on perturbative BPS-states that exists in a type II 
superstring where the space-dimensions are compactified on
$S_1\times C_4$, and $C_4$ is some Calabi-Yau
fourfold. This construction is analogous to a
construction in \cite{BPS}. A general vertex
operator in this theory can be written as
\be e^{k_L \cdot X_L(z)} e^{k_R \cdot X_R(\bar{z})}
    V(z,\bar{z}) \ee
where
\be k_L = (E,P_L), \hspace{1cm} k_R = (E,P_R) \ee
and $P_L,P_R$ are the left- and right moving momenta
in the $S_1$ direction. They are parametrized by the
Narain-lattice $\Gamma^{1,1}$, and can be expressed
in terms of winding and momentum numbers $m,n$ as follows
\be P_L = \frac{m}{R}-\frac{nR}{2}, \hspace{1cm}
    P_R = \frac{m}{R}+\frac{nR}{2} \ee
Level matching implies that $V$ has left and right
conformal dimensions (including ghosts) given by
$(n_L,n_R) = (\frac{1}{2}k_L^2,\frac{1}{2}k_R^2)$.
The BPS-condition is simply $n_R=0$. implying
$E^2=P_R^2$, and then
\be n_L = \frac{1}{2}k_L^2 = \frac{1}{2}(P_R^2-P_L^2)
  = mn \ee
I choose the solution $E=P_R$, leading to the BPS-states.
(The other solution gives anti-BPS states).
The product of two BPS-states is defined in \cite{BPS} as
\be {\cal R}(V_1,V_2)=P(\lim_{z_1\rightarrow z_2}
  V_(z_1,\bar{z}_1) V_2(z_2,\bar{z}_2) \bmod Q) \ee
where $P$ projects the resulting vertex operator
to the appropriate ghost-numbers (using the operator
$b_0 \otimes \bar{b}_0$). In the case that the vertex-operators
can be factorized into left- and right moving components
(up to a cocycle factor), say
\be V_k(z,\bar{z})=V_k(z) \otimes \bar{V}(\bar{z}) \epsilon_k \ee
then using~(\ref{dotbrak}) it follows that the product
can be written as
\be {\cal R}(V_1,V_2) = \{ V_1, V_2 \} \otimes
      \{ \bar{V}_1, \bar{V}_2 \} \epsilon_1\epsilon_2 \ee
So it has the structure of a tensor product of two 
Lie-algebras, and the algebra on the left-movers
separately is a Lie-algebra, which by the BPS-condition,
is graded by $\Gamma^{1,1}$.
Note that this product closes on the space of BPS-states,
without the use of Lorentz-boosts. This is by virtue of the
fact that the right moving momenta of all BPS-states
have the same (lightlike) direction, with a spatial component
only in the $S_1$ direction, and none in the 
$C_4$ direction (by assumption).
This is so because momentum-operators are ill-defined on a
general $C_4$. If, however, the Calabi-Yau $C_4$ is such
that momentum-operators can be defined,
(for example on a torus) then my definition
of the BPS-algebra defines a subalgebra of the usual BPS-algebra,
a subalgebra with a specific choice for the right-moving
momentum. The entire algebra in this case can only close
using Lorentz-boosts of \cite{BPS}. 

In the case of toroidal compactification, the algebra
on the left-moving parts of the BPS-vertex operators (with
given right-moving momentum direction) is just the GKM-algebra
as considered in the previous section. It is a subalgebra
of the type II equivalent of 
${\cal H}_0^{\mbox{\scriptsize mult}}$ in \cite{BPS}.

\subsection{Half-twisted model.}

In this section I define a Lie-algebra, associated
to a Calabi-Yau fourfold $C_4$, which has the same
graded dimensions as the BPS-algebra defined in the
previous section. This construction is based on a
so called half-twisted model of \cite{halftwist}.
The claim is that the non-linear sigma model
associated to $C_4$, with the right-moving modes
restricted to the NS-sector, has the structure of a
$\bz_2\times \bz_2$-graded vertex algebra modulo
$\bar{Q}$, where $\bar{Q}$ is the twisted
right-moving BRST-operator, equal to 
$\bar{G}^+_{-\frac{1}{2}}$. Indeed, the OPE
\be V_1(z_1,\bar{z}) V_2(z_2,\bar{z}) =
  \sum_n \frac{W_n(z_2,\bar{z})}{(z_1-z_2)^n} \bmod \bar{Q} \ee
is well defined, since there are no poles in $\bar{z}$
\cite{chiral}. In the case that the vertex-operators can be
factorized into left- and right moving components
(again up to a cocycle factor), it is not hard to see
that this vertex algebra is the tensor product of
the vertex algebra on the left-moving modes, and the
associative algebra formed by the right moving chiral-ring.
The supersymmetric index of this product vertex algebra
is equal to the elliptic genus of $C_4$. If this algebra
is used as transverse algebra in the construction
as was described in section~\ref{sch}, the result is
a unique Lie-algebra associated to the Calabi-Yau fourfold.
In the factorizable case, this Lie-algebra is the tensor
product of a Lie-algebra on left-moving modes with
an associative algebra on the right moving modes.

\section{Orbifold constructions.}
\setcounter{equation}{0}

In this section I will sketch the orbifold compactification
of the superstring on $E_{1,1}\otimes (E_8/\bz_2)$.
The vertex-algebra is constructed by considering
a $\bz_2$-twisted construction of the transverse
part of the algebra, very much like the construction
of the monster vertex algebra \cite{FLM}.
The orbifold is defined by $\theta: V^s\rightarrow V^s$,
$\theta^2=1$, which acts as the identity on $V^L_{\gamma}$,
and
\be \theta V(D_4^*)_{\gamma} \rightarrow
 \left\{ \begin{array}{rl}
 V(D_4^*)_{\gamma} & \gamma=00,11 \\
 -V(D_4^*)_{\gamma} & \gamma=10,01 \end{array}
 \right. \ee
\be \theta V(E_8) \rightarrow V(-E_8) \ee
Note that $\theta$ also
commutes with $Q$. So the cohomology can be split
into the positive and negative eigenspaces of
$\theta$. I'm interested in the dimensions of these
eigenspaces. Since the dimensions are described
in terms of the dimensions of the transverse algebra,
I need to know how this space splits. 
The relevant formulas for $V(E_8)$ can be found
in \cite{FLM}. It splits into spaces with
partition functions
\be B^{\pm} = \frac{1}{2} \left(
 \frac{E_4(q)}{\eta^8(q)} 
 \pm \frac{\eta^8(q)}{\eta^8(q^2)} \right) \ee
And so the spaces $V_{01}^T$ and $V_{11}^T$, relevant
for the cohomology after GSO-projection, split into
spaces with (extended) partition functions
\be \mbox{Tr}_{(V^T_{\gamma})^{\pm}} 
   q^{L_0-1/2} y^{J_0} = \left\{ \begin{array}{ll}
  B^- \theta_2   & \gamma=01,(+) \\
  B^+ \theta_2   & \gamma=01,(-) \\
  B^+ \theta^-_1 & \gamma=11,(+) \\
  B^- \theta^-_1 & \gamma=11,(-) \end{array}
  \right. \ee
And the supersymmetric index of the untwisted
transverse part splits into
\be I^+ = B^+ \theta^-_1 - B^- \theta_2 \ee
\be I^- = B^- \theta^-_1 - B^+ \theta_2 \ee
Next, the twisted transverse part
can be constructed out of twisted versions of
$V(E_8)$ and $V(D_4^*)$. However, twisting of
$V(D_4^*)$ (which itself
can be considered as the direct sum of
$V(\bz^4)$ and a $\theta$-twisted module of it
\cite{spinor}), gives back the space $V(D_4^*)$.
The twisting of $V(E_8)$ gives rise to
a space $V^T(E_8)$, and the action of
$\theta$ can be extended in a natural way
to this space such that it splits into spaces
with partition functions (see again \cite{FLM})
\be B_T^{\pm} = 8\left(
 \frac{\eta^8(q)}{\eta^8(\sqrt{q})} \pm
 \frac{\eta^8(q^2)\eta^8(\sqrt{q})}{\eta^{16}(q)} \right) \ee
The factor $8$ comes from the fact that
the lattice algebra of $E_8$ in the twisted
sector has a unique $16$-dimensional representation.
Note the strange identity $B^-=B_T^-$.
So the twisted sector in the orbifold construction
is made by replacing the transverse algebra
with a twisted algebra as follows
\be V(E_8)\otimes V_{\gamma} \rightarrow
       V^T(E_8)\otimes V_{\gamma+11} \ee
(where the $V_{\gamma}$ are part of $V(D_4^*)$).
The twisted components relevant for the
cohomology in the twisted sector, call them
$(V_{01}^T)_T$ and $(V_{11}^T)_T$, split
under the extended action of $\theta$, and
\be \mbox{Tr}_{(V^T_{\gamma})_T^{\pm}}
   q^{L_0-1/2} y^{J_0} = \left\{ \begin{array}{ll}
  B_T^- \theta_2 & \gamma=01,(+) \\
  B_T^+ \theta_2 & \gamma=01,(-) \\
  B_T^+ \theta^+_1 & \gamma=11,(+) \\
  B_T^- \theta^+_1 & \gamma=11,(-) \end{array}
  \right. \ee
with related supersymmetric indices for the
twisted sector
\be I^+_T = B_T^+ \theta^+_1 - B_T^- \theta_2 \ee 
\be I^-_T = B_T^- \theta^+_1 - B_T^+ \theta_2 \ee
The entire GSO-projected orbifold-algebra falls apart
into four spaces, which I will denote by
\be W_{00}=(V_{00}\oplus V_{10})^+, \hspace{1cm}
    W_{01}=(V_{00}\oplus V_{10})^-  \ee
\[  W_{10}=(V_{00}\oplus V_{10})_T^+, \hspace{1cm}
    W_{11}=(V_{00}\oplus V_{10})_T^- \]
This algebra is very much like the algebra
structure described by Huang in \cite{related}.
It's a $\bz_2\times\bz_2$-graded vertex algebra.
In particular, the space $W_{00}\oplus W_{10}$
can be given the structure of a vertex superalgebra.
This algebra corresponds to a chiral orbifold
of the superstring. Its construction is very similar
to the construction of the monster vertex-algebra
\cite{FLM}. Note that the supersymmetric index
associated with the transverse algebra has the
properties of a weak Jacobi form of index $2$ and
weight $0$, and actually
\be I^+ + I_T^+ = \frac{1}{6}(5J_1+J_2) = 
  (y^{-2}+22+y^2) +O(q) \ee
(By the way, the function $I_T^+-I^-=\frac{1}{6}
(J_2-J_1)$ is proportional to the function
that defines the denominator formula for a
generalized Kac-Moody algebra which is an automorphic
form correction to the Kac-Moody algebra defined
by the symmetrized generalized Cartan matrix
\be G_2=\left( \begin{array}{cccc}
4&-4&-12&-4 \\ -4&4&-4&-12 \\
-12&-4&4&-4 \\ -4&-12&-4&4 \end{array} \right) \ee
as was found in \cite{grit}. This might suggest
relations between this chiral orbifold construction
and their GKM-algebra.)
Instead of looking at a chiral orbifold, I can look
at the orbifold of the closed superstring. As is well
known, this one is described in terms of the space
\be \bigoplus_{\gamma\in\bz_2\times\bz_2}
   W^L_{\gamma} \oplus W^R_{\gamma} \ee
where $W^{L,R}_{\gamma}$ are two copies (left- and 
right-moving) of $W_{\gamma}$. Now again I replace
the right-moving algebra by the associative algebra
of supersymmetric groundstates $C_{\gamma}$, which like
$W_{\gamma}$ is $\bz_2\times\bz_2$-graded. Note that
$C_{11}$ has dimension zero, and
\be 8I^+(q,y)-8I^-(q,y)+16I_T^+(q,y)+0I_T^-(q,y) \ee
\[ = \frac{1}{3}(16J_1+8J_2)=
  (8y^{-2}+32y^{-1}+304+32y+8y^2) +O(q) \]
correctly reproduces the elliptic genus of the
orbifold $T_8/\bz_2$, which is a singular manifold
with orbifold Euler number $384$.
The prefactors are of course the
superdimensions of the respective $C_{\gamma}$.

\vspace{1\baselineskip} \noindent
{\bf Acknowledgement.}

\noindent I would like to thank
R. Dijkgraaf for helpful discussions.

\end{document}